\documentclass[aps,preprint,superscriptaddress,floatfix,showpacs,amsmath,amssymb,nofootinbib]{revtex4}

\usepackage{graphicx}% Include figure files
\usepackage{dcolumn}% Align table columns on decimal point
\usepackage{bm}% bold math
\usepackage{color}
\usepackage{tabularx}
\usepackage{csquotes}
\usepackage[normalem]{ulem}

\newcommand{\ncto}{Na$_2$Co$_2$TeO$_6$}

\newcommand{\tc}{$T_{\rm cr}$}
\newcommand{\tn}{$T_{\rm N}$}

\newcommand{\ac}{$\alpha_{\rm c}$}
\newcommand{\cptot}{$c_{\rm p}$}

\newcommand{\bc}{$B_\mathrm{C}$}

\newcommand{\beginsupplement}{%
        \setcounter{table}{0}
        \renewcommand{\thetable}{S\arabic{table}}%
        \setcounter{figure}{0}
        \renewcommand{\thefigure}{S\arabic{figure}}%
     }

\begin{document}

%\preprint{APS/123-QED}

\title{Signatures of a Quantum Critical Endpoint in the Kitaev Candidate Na$_2$Co$_2$TeO$_6$}

%%%%  AUTHORS %%%%%%%%%%%%%%%
\author{J.~Arneth}\email{jan.arneth@kip.uni-heidelberg.de}
\affiliation{Kirchhoff Institute for Physics, Heidelberg University, INF 227, D-69120 Heidelberg, Germany}
\author{K.-Y.~Choi}
\affiliation{Department of Physics, Sungkyunkwan University, Suwon 16419, Republic of Korea}
\author{R. Kalaivanan}\affiliation{Institute of Physics, Academia Sinica, Taipei 11529, Taiwan}
\author{R. Sankar}\affiliation{Institute of Physics, Academia Sinica, Taipei 11529, Taiwan}
\author{R.~Klingeler}\email{klingeler@kip.uni-heidelberg.de}
\affiliation{Kirchhoff Institute for Physics, Heidelberg University, INF 227, D-69120 Heidelberg, Germany}

\date{\today}

\begin{abstract}
The putative Kitaev material \ncto\ has recently been proposed to enter a quantum spin disordered state when magnetic fields are applied in parallel to the honeycomb layers. In this report we uncover signatures of a quantum critical endpoint (QCEP) associated with the assumed order-disorder transition by means of high-resolution capacitance dilatometry. At the critical field \bc\ $\simeq 6$~T , a sign change of the out-of-plane thermal expansion coefficient \ac\ indicates accumulation of entropy upon crossing the phase boundary. The proportional relationship between isothermal magnetisation and magnetostriction signals that the QCEP can be tuned by magnetic field and pressure simultaneously. The presented results expand the material classes that exhibit metamagnetic quantum criticality to honeycomb antiferromagnets with possible Kitaev interactions.
\end{abstract}
\maketitle

Within the framework of the everlasting search for a quantum spin liquid (QSL), the exactly solvable Kitaev model offers a promising route to a long-range entangled yet magnetically disordered state with exotic elementary excitations, such as Majorana fermions~\cite{kitaev2006}. At the heart of this model lie strongly anisotropic, bond-dependent Ising interactions in a $S = 1/2$ honeycomb network, leading to pronounced magnetic frustration. The first wave of Kitaev candidate materials consisted of $d^5$ transition metal ions with strong spin-orbit coupling (SOC) situated in edge-sharing anion octahedra~\cite{jackeli2009,takagi2019}. Among many experimental results arising from such theoretical proposals especially the $5d$ iridate Na$_2$IrO$_3$ and the $4d$ ruthenate $\alpha$-RuCl$_3$ have managed to stand out, since both exhibit promising characteristics hinting at proximity to the QSL phase~\cite{chun2015,do2017,kasahara2018}.

However, over time the list of materials possibly realizing the Kitaev model has also expanded to include cobaltates with high-spin $d^7$ electronic configuration~\cite{liu2018_ncto,sano2018,liu2020}. In these systems the presence of spin-active $e_g$ electrons is suggested to reduce the strength of Heisenberg interactions $J$, thereby enhancing the dominance of the Kitaev contributions. Out of these "second wave" materials, \ncto\ has quickly been regarded as the most promising compound due to its notable similarities to $\alpha$-RuCl$_3$: Even though \ncto\ exhibits magnetic long-range order below $T_\mathrm{N} = 27\,\mathrm{K}$, the firstly reported antiferromagnetic zig-zag ground state~\cite{bera2017} -- recently revised as a triple-\textbf{q} order~\cite{chen2021_ncto} -- is a promising indication of a proximate QSL state, as these two phases are adjacent in the parameter space of the Kitaev-Heisenberg model~\cite{chaloupka2013}. Indeed, a discontinuous phase transition appears when a magnetic field of $B_\mathrm{C} \sim 6\,\mathrm{T}$ is applied along the direction of the nearest-neighbour Co-Co bonds~\cite{xiao2019,yao2020}. Despite the lack of knowledge about the exact magnetic ground state in the high-field phase of \ncto, the experimental and theoretical results suggest the emergence of magnetic disorder at the critical field~\cite{lin2021,xiang2023,hong2024}.

Here, we further highlight the intriguing effects of in-plane magnetic fields on \ncto\ by showing that the first-order phase transition at \bc\ exhibits strong signatures of quantum critical behaviour close to a quantum critical endpoint (QCEP). The role of magnetoelastic coupling is elucidated by thermal expansion and magnetostriction measurements between 2~K and 300~K and up to 15~T on high-quality single crystals of \ncto\ grown as described in Ref.~\onlinecite{lee2021}. Note, that the presence of a potential antiferromagnetic impurity phase~\cite{dufault2023} is ruled out by our magnetisation data (cf. Supplemental Material Ref.~\onlinecite{supplement}). Out-of-plane relative length changes $dL_\mathrm{c}/L_\mathrm{c}$, i.e., along the crystallographic $c$-direction, were studied on a 0.145~mm thin single crystal by means of a three-terminal high-resolution capacitance dilatometer (Kuechler Innovative Measurement Technology) in a home-built set-up placed inside a Variable Temperature Insert of an Oxford magnet system~\cite{Kuechler2012, Werner2017}. From the relative length changes the linear thermal expansion coefficient $\alpha_{\rm c} = 1/L_{\rm c}\times(\partial L_{\rm c}/\partial T$) was derived. Magnetic fields were applied along the direction of first-neighbour Co-Co bonds ($B||a^*$) within the hexagonal planes. Thermal expansion and magnetostriction measurements were performed at rates of $0.3~$K/min and $0.3~$T/min, respectively. The magnetisation was studied by means of a Physical Properties Measurement System (PPMS, Quantum Design) using the vibrating sample magnetometer (VSM) option.

\begin{figure}[th]
    \includegraphics[width = 0.8\columnwidth]{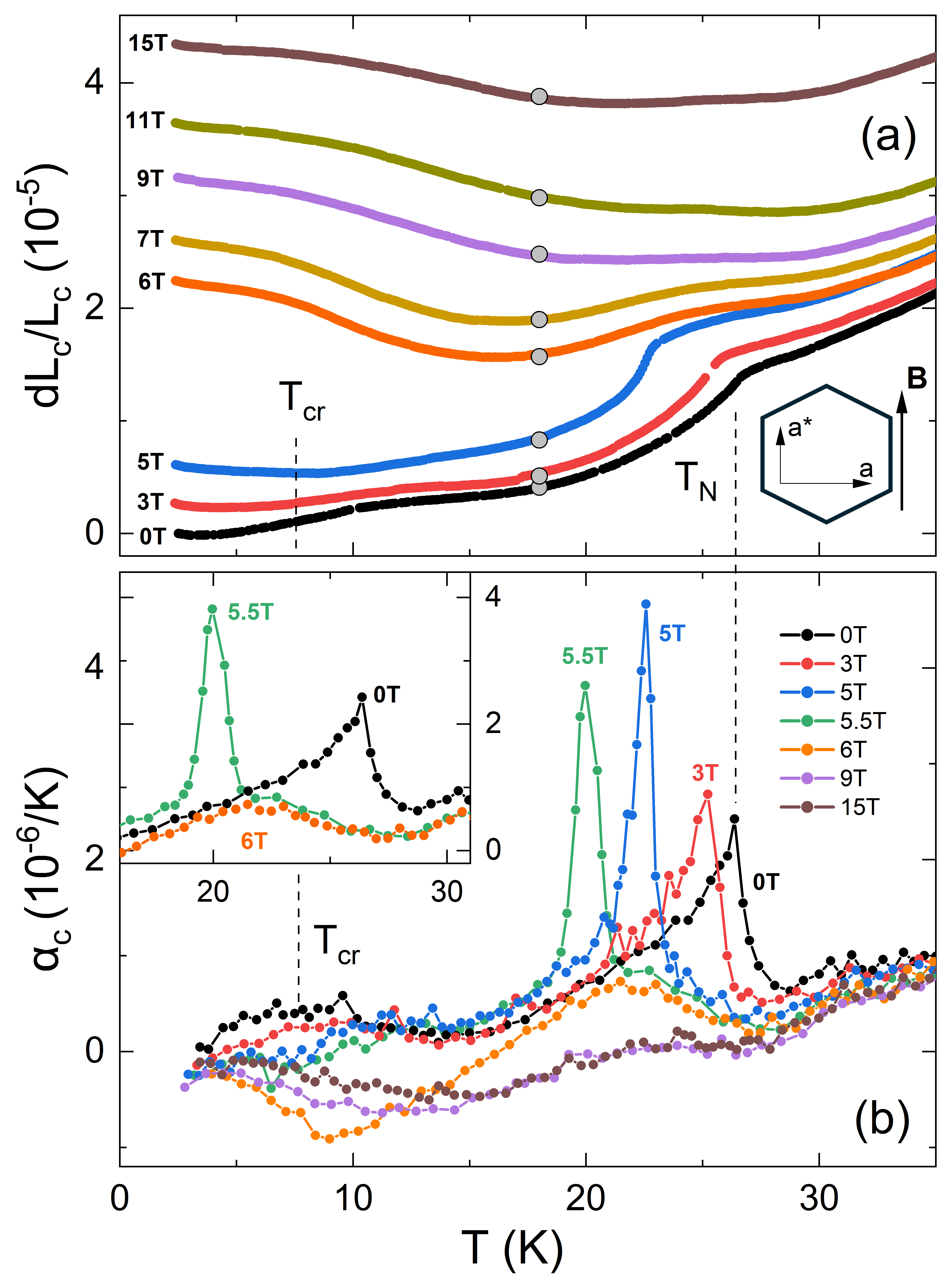}
    \caption{(a) Out-of-plane thermal expansion $dL_\mathrm{c}/L_\mathrm{c}$ and (b) linear thermal expansion coefficient $\alpha_\mathrm{c}$ for both zero and finite magnetic fields applied $B||a^*$. As a reference point, $L(T=2\,\mathrm{K};B=0\,\mathrm{T})$ is set to zero and data at different fields in (a) are scaled according to the magnetostriction measurements at 18~K (grey markers; see Fig.~\ref{fig:MS}). \tn\ and \tc\ mark the evolution of long-range magnetic order and the additional hump in zero field. Inset: Anomalies in $\alpha_\mathrm{c}$ at \tn\ for $B=0$, 5.5, and 6~T.}
    \label{fig:TE}
\end{figure}

The low-temperature out-of-plane thermal expansion of \ncto\ is shown in Fig.~\ref{fig:TE} for representative magnetic fields $B||a^*$.\footnote{The measurements have been performed following a field-cooled (FC) protocol where the sample was cooled in the applied field.}
In zero magnetic field, the $c$ axis continuously shrinks upon cooling, but notably, it also shows a sharp kink at the antiferromagnetic ordering temperature $T_{\rm N} = 26.8(3)\,\mathrm{K}$, which is reflected by a corresponding $\lambda$-shaped peak in the thermal expansion coefficient \ac\ (Fig.~\ref{fig:TE}b). The pronounced anomaly unambiguously indicates significant coupling between spin and lattice degrees of freedom in \ncto. Interestingly, at around $T_\mathrm{cr} = 7.5(9)$~K, another step-like feature can be found in the thermal expansion data as well as a corresponding broad hump in \ac . Note, that the observation of a $\lambda$-shaped anomaly at \tn\ and a broad hump at \tc\ resembles the overall behaviour of the specific heat capacity \cptot\ of \ncto\ as reported in Ref.~\onlinecite{yao2020}. In this publication, the  broad hump at \tc\ is attributed to magnetic inter-plane correlations that remain short-ranged due to Na-disorder in separating layers. 

While the thermal expansion coefficient at temperatures above \tn\ is not particularly sensitive to magnetic fields applied $|| a^*$ axis, the anomaly at \tn\ shifts to lower temperatures and becomes more pronounced in magnetic fields up to 5~T. Concomitantly, the initially $\lambda$-shaped anomaly evolves into an even more pronounced symmetric peak and a small preceding step-like feature (see \ac (5~T) in the inset in Fig.~\ref{fig:TE}b). The observed symmetric peak-shaped anomaly implies an associated jump in $L_{\rm c}$, i.e., a first-order phase transition. At fields $B \geq 6\,\mathrm{T}$, no clear anomaly is visible in the length changes anymore and in \ac\ only the small step-like increase can be identified. Also for the hump-like feature at \tc , two field regimes can be identified: For $B \leq 5\,\mathrm{T}$, the step-like shrinking of the $c$ axis continuously diminishes with increasing external magnetic field while it reappears for $B\leq 6$~T, albeit now signaling an anomalous {\it increase} in length (Fig.~\ref{fig:TE}a). Eventually, at $B>10$~T, the $c$ axis only weakly depends on temperature $T<30$~K and \ac\ is independent on the applied field in this regime (see the Supplemental Material Ref.~\onlinecite{supplement} for all thermal expansion data). For the further analysis of the data, we hence use \ac (15~T) as an estimate of the non-magnetic thermal expansion, since at such high fields the spins are completely polarized at sufficiently low temperatures. Similar methods have been used in the literature, e.g., in the closely related honeycomb magnet $\alpha$-RuCl$_3$~\cite{gass2020}.

\begin{figure}[t]
    \includegraphics[width = 0.8\columnwidth]{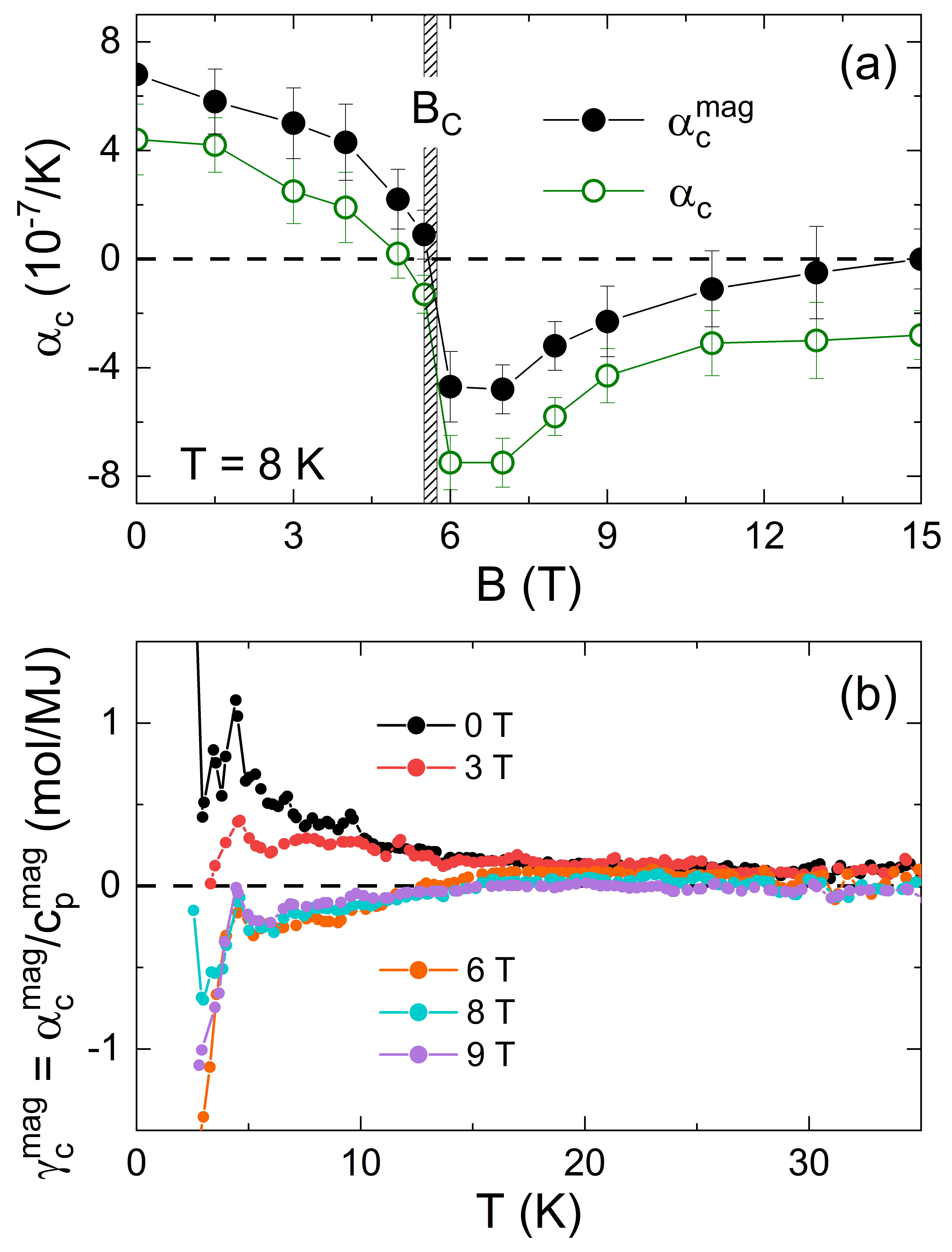}
    \caption{(a) Out-of-plane linear thermal expansion coefficient and its magnetic contribution at 8~K as a function of magnetic field. The vertical dashed region marks the first-order phase transition at \bc . (b) Temperature dependence of the unidirectional magnetic Grüneisen parameter $\gamma_c^{\rm mag}$, obtained by dividing the magnetic contributions to $\alpha_{c}$ and $c_{\rm p}$ from Ref.~\onlinecite{yao2020}, at selected magnetic fields applied $B||a^*$.}
    \label{fig:grueneisen}
\end{figure}

We particularly emphasize the field-dependence of the thermal expansion around \tc : It is already evident  from the $c$ axis dependence in Fig.~\ref{fig:TE}a that \ac\ shows a sign change from positive at small fields to negative at fields $B \geq 6\,\mathrm{T}$. At intermediate fields around 5.5~T the thermal expansion coefficient almost vanishes below 10~K.
The field effects are particularly visible when considering the thermal expansion coefficient as a function of $B$ at fixed temperature of $T=8$~K
as shown in Fig.~\ref{fig:grueneisen}a (for other temperatures see the Supplemental Material Ref.~\onlinecite{supplement}).\footnote{At $T=8$~K, zero-field thermal expansion shows a hump associated to pronounced critical fluctuations so that the effects of magnetic fields can be expected to be particularly large. In addition, at lowest temperatures, thermal expansion vanishes so that the signal-to-noise ratio decreases at lowest temperatures under study.}
Here, we also plot the bare magnetic contribution $\alpha^{\rm mag}_{\rm c}=\alpha_{\rm c}-\alpha^{\rm non-mag}_{\rm c}$ by using $\alpha^{\rm non-mag}_{\rm c}\simeq \alpha_{\rm c}(15~{\rm T})$. Both datasets clearly imply a sign change in thermal expansion at $B_{\rm C}(8~{\rm K})\simeq 5.5$~T which coincides well with the field-driven first-order phase transition occurring in the same field-range~\cite{viciu2007,lefrancois2016,yao2020,lin2021}.

In recent literature, the corresponding discontinuity in the magnetisation at \bc\  %$B_\mathrm{C} = 6.1\,\mathrm{T}$ at 2~K 
has been attributed to an order-disorder transition into a QSL-like state~\cite{lin2021}. Here, we uncover another intriguing property of the anomalous behaviour at \bc, that is, its distinct signatures of quantum criticality. Since the linear thermal expansion coefficient is directly proportional to the pressure dependence of the entropy via the Maxwell relation $\alpha_\mathrm{c} \sim \partial S/\partial p_\mathrm{c}|_{T}$, a sign change of \ac\ upon the application of magnetic fields implies a maximum of the magnetic contribution to the entropy at \bc . Such accumulation of $S_\mathrm{mag}$ is regularly expected to occur at a quantum critical point (QCP)~\cite{garst2005,millis2002, gegenwart2016}. It is important to note, that the here reported quantum criticality, and hence also its critical properties, differs from the commonly discussed ones in the way that the underlying phase transition is not continuous but of a first-order type. In this scenario the so-called metamagnetic quantum critical endpoint (QCEP) is understood to be the endpoint of a line of a first-order phase transition when the temperature is reduced to zero~\cite{garst2005,millis2002,beneke2021}. Our interpretation is further supported by the data in Ref.~\onlinecite{zhang2023} the detailed inspection of which shows that the in-plane thermal expansion coefficient $\alpha_{\rm a^*}$ changes its sign, too, at $B||a^* = 6$~T~\cite{zhang2023}.

A key signature of metamagnetic quantum criticality is a sign change of the magnetic Grüneisen parameter, which is defined as the ratio of the magnetic contributions to the thermal expansion coefficient and the specific heat capacity $\gamma_\mathrm{mag} = \alpha^\mathrm{mag}/c^\mathrm{mag}_p$~\cite{garst2005, gegenwart2006, millis2002}. Figure~\ref{fig:grueneisen}b shows the uniaxial magnetic Grüneisen parameter of \ncto\ at representative magnetic fields as calculated from our thermal expansion data and the specific heat capacity measured in Ref.~\onlinecite{yao2020}. The resulting Grüneisen ratio is close to zero at $T>20$~K but continuously increases upon cooling for $B\leq 5$~T. The opposite trend appears for $B\geq 6$~T. Extrapolating the temperature dependence for $T \rightarrow 0$ the data indicate a divergence at zero temperature, which can, however, not be seen directly in the data due to limited resolution and tiny length changes at low temperatures.

Signatures of metamagnetic quantum criticality as described in the present study have been first found and discussed in Sr$_3$Ru$_2$O$_7$~\cite{grigera2001, grigera2004, gegenwart2006}. Later on, a sign change of the out-of-plane thermal expansion coefficient connected to a first-order phase transition was also observed in CeRu$_2$Si$_2$~\cite{paulsen1990, weickert2010} and in Ca$_{1.8}$Sr$_{0.2}$RuO$_4$~\cite{baier2007}. Additionally, indications of QCEPs were found in UGe$_2$~\cite{taufour2010}, UCoAl~\cite{aoki2011} and ZrZn$_2$~\cite{uhlarz2004}. To our knowledge, metamagnetic quantum criticality has been reported exclusively in itinerant electron systems, which would render \ncto\ the first antiferromagnetic insulator to exhibit a quantum critical endpoint.
At this point it is noteworthy that thermal expansion in the closely related Kitaev candidate $\alpha$-RuCl$_3$ also hints at the emergence of a quantum phase transition at \bc, which is debated to precede a QSL state~\cite{gass2020}. In $\alpha$-RuCl$_3$, however, the transition into the putative magnetically disordered phase exhibits a rather continuous character, which leads to differing quantum critical properties as compared to \ncto.

\begin{figure}[t]
    \centering
    \includegraphics[width = \columnwidth]{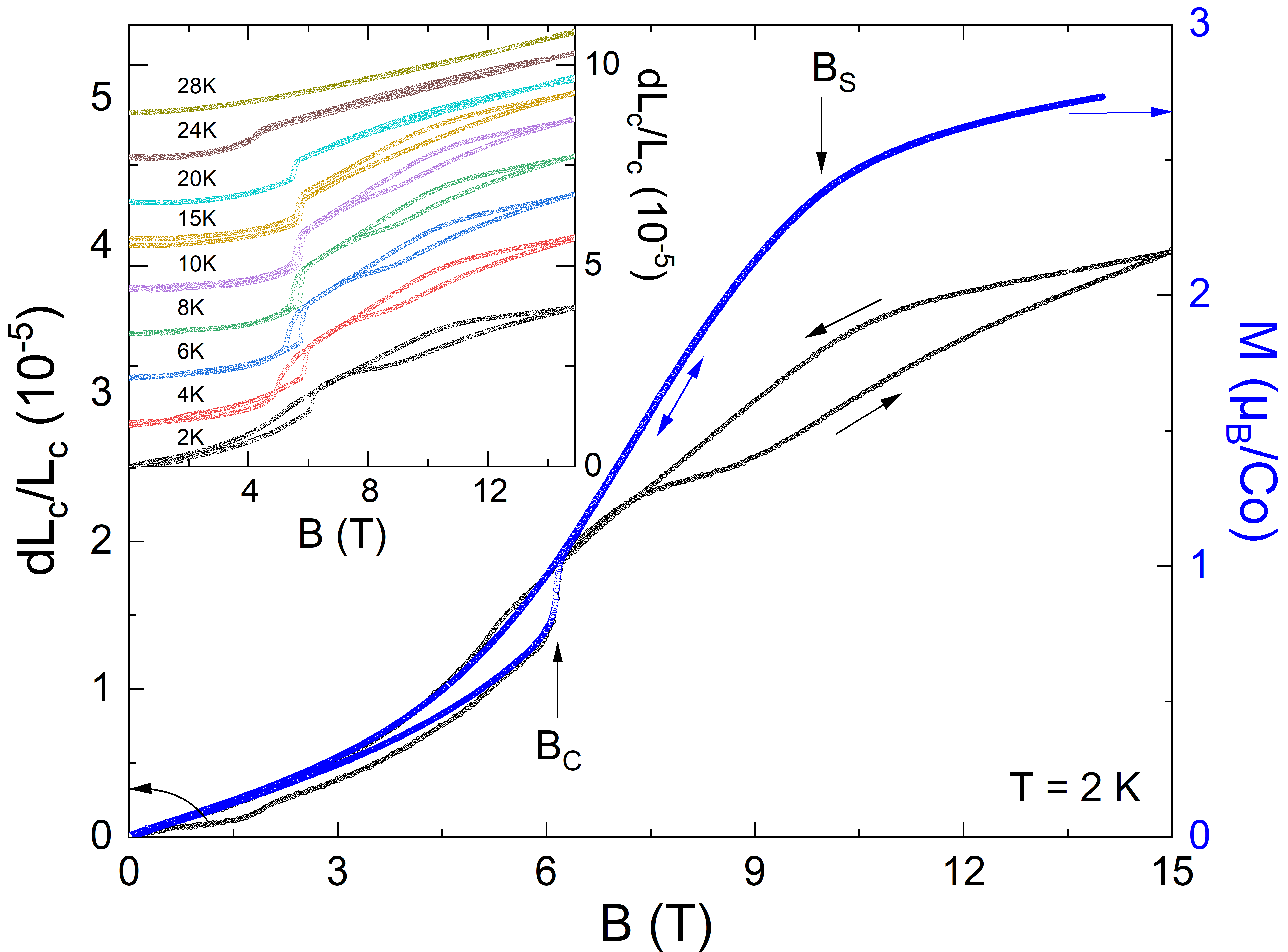}
    \caption{Isothermal magnetostriction $L_{\rm c}(B)$  (black) and magnetisation $M(B)$ (blue) at $T = 2$~K for $B||a^*$. Arrows denote the direction of the field-sweep. The inset depicts magnetostriction measurements at various temperatures, which are shifted along the ordinate. $B_\mathrm{C}$ and $B_\mathrm{S}$ mark the critical field and the saturation field, respectively.}
    \label{fig:MS}
\end{figure}

The coupling of spin and lattice degrees of freedom can be also directly studied through magnetostriction experiments. Figure~\ref{fig:MS} shows the relative out-of-plane length changes (black) and the isothermal magnetisation (blue) of \ncto\ at 1.8~K for $B||a^*$. As can be seen, the discontinuity in the magnetisation is accompanied by a positive jump in the $c$ axis length, suggesting that both magnetic field and uniaxial pressure can be used as tuning parameters of the QCEP~\cite{garst2005}. At $B_{\rm S}(2~{\rm K})\simeq 10$~T, the magnetisation implies the evolution of the polarised spin configuration.
Similar to the field-region around \bc, $dL_\mathrm{c}$ exhibits pronounced differences between the up- and down-sweep of the magnetic field starting from 8~T, while no corresponding hysteresis is observed in the magnetisation data. 
As depicted in the inset of Fig.~\ref{fig:MS}, the hysteresis regions at \bc ($T$) narrows as the temperature is increased and totally vanishes at $T \geq T_\mathrm{N}$. Notably, hysteresis in magnetostriction above \bc ($T$) ranges well beyond $B_{\rm S}(T)$ and persists up to at least 24~K, i.e., in the whole reported quantum disordered regime~\cite{lin2021}. The hysteresis diminishes but the hysteretic field range does not shrink upon heating. This rather indicates magnetostructural domain effects than a weak first order character of the phase boundary $B_{\rm S}(T)$. In addition, the magnetostriction exhibits a weak discontinuity of unknown origin at $B \simeq 1.5$~T when the field is increased from zero, but not when the field is subsequently decreased. At intermediate temperatures around $6$~K this results in crossing of the up- and down-sweep magnetostriction at approximately $5.5$~T.

Noteworthy, field-induced magnetisation and length changes, at 2~K, closely track each other, i.e., the curves can be well overlapped by a single scaling factor, up to the critical field \bc . This observation parallels the findings of magnetostriction measurements on Sr$_3$Ru$_2$O$_7$~\cite{grigera2004} and CeRu$_2$Si$_2$~\cite{weickert2010}. Following the arguments discussed in that frame, such a good proportionality of $M(B)$ and $L_\mathrm{c}(B)$ implies that variations in uniaxial pressure and magnetic field probe the same thermodynamic information as expected in a quantum critical regime~\cite{garst2005}. Hence, the magnetostriction data strongly corroborate the results of our thermal expansion measurements.

\begin{figure}[]
    \centering
    \includegraphics[width = \columnwidth]{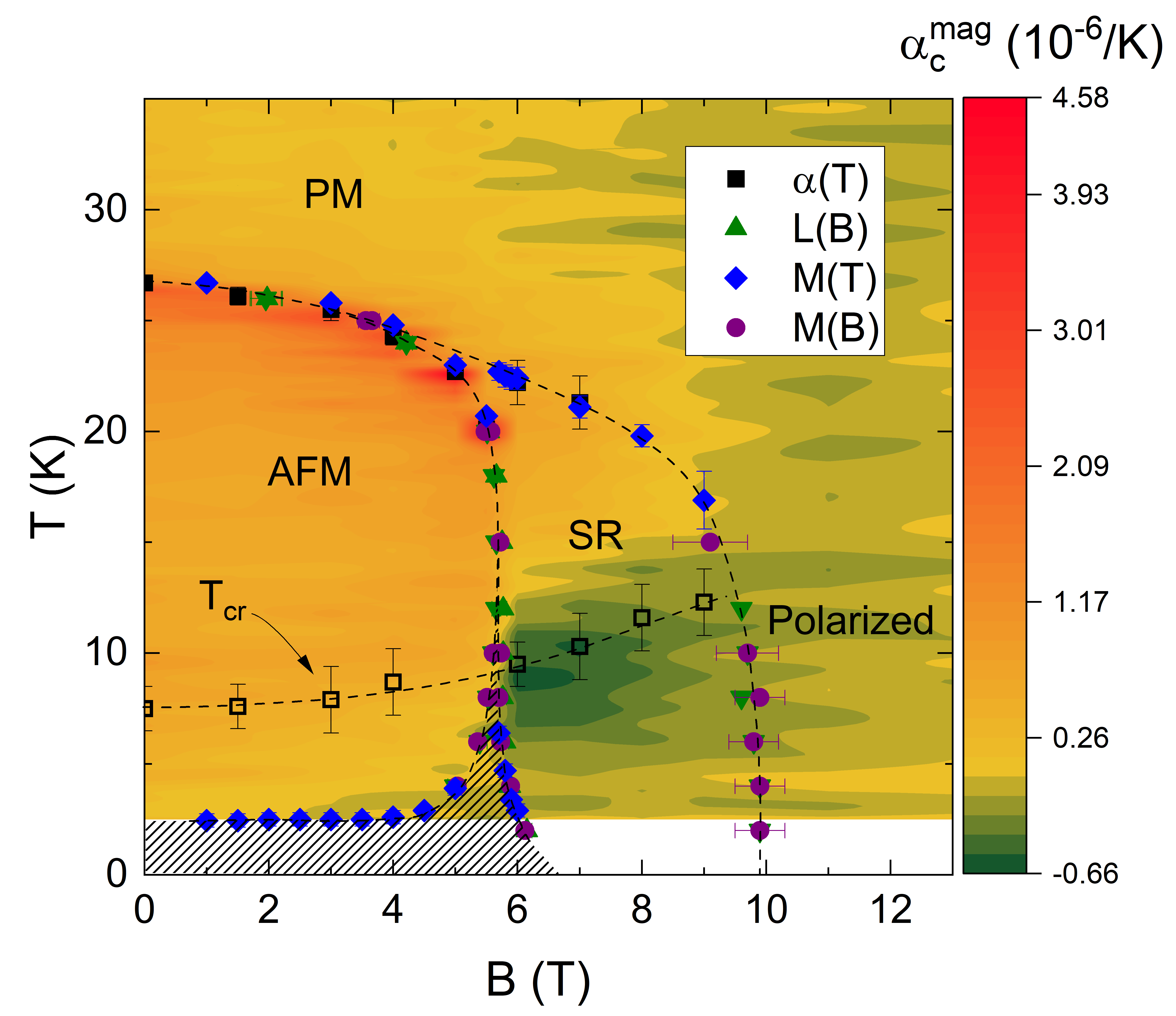}
    \caption{Magnetic phase diagram of \ncto\ for magnetic fields applied $B||a^*$ as constructed from magnetic and dilatometric measurements. PM, AFM and SR denote the paramagnetic, canted antiferromagnetic and spin reoriented phase, respectively. The hatched area shows the hysteresis region observed in $M$ and $L_{\rm c}$ (see the text). Additionally, $\alpha^ \mathrm{mag}_\mathrm{c}(T,B)$ is depicted in colour scale; The distinct data sets are shown in the Supplemental Material Ref.~\onlinecite{supplement}.}
    \label{fig:PhD}
\end{figure}

Collecting the anomalies found in dilatometric and magnetic quantities allows for the construction of the magnetoelastic phase diagram shown in Fig.~\ref{fig:PhD} (see the Supplemental Material Ref.~\onlinecite{supplement} for the magnetic measurements). In agreement with most phase diagrams reported in the literature~\cite{lin2021,yao2020}, our dilatometric studies imply three distinct thermodynamic phases: Coming from the high-temperature paramagnetic (PM) regime the spin system undergoes a continuous phase transition into a canted antiferromagnetic (AFM) state at \tn. Upon increasing the magnetic field along the Co-Co bonds, \ncto\ exhibits a first-order phase transition showing strong signs of quantum criticality. Above \bc , the spins reorient (SR) out of their low-field configuration into a yet unknown magnetic structure while, above $B_{\rm s}(2~{\rm K})=10$~T, the magnetic moments are almost fully polarized. As indicated by the shaded area in Fig.~\ref{fig:PhD}, the AFM/SR phase boundary exhibits pronounced hysteresis at low temperatures. At low temperatures, our data suggests that the system does not return into the AFM phase even when the magnetic field is decreased back to zero.

In conclusion, our measurements of the thermal expansion and magnetostricition consequently give insights into the important role of magnetoelastic coupling in the Kitaev candidate material \ncto. A sign change of the Grüneisen parameter at the critical field \bc\ indicates the accumulation of entropy in the ($B$,$T$)-plane above the critical field and thereby implies the existence of a metamagnetic QCEP. In addition, the proportional relationship between isothermal magnetisation and magnetostriction for $B\leq B_{\rm c}$ signals that the QCEP can be tuned by magnetic field and pressure simultaneously. The presented results expand the scope of sample classes that exhibit metamagnetic quantum criticality to honeycomb antiferromagnets with possibly Kitaev-like interactions and, hence, contribute to the understanding of quantum criticality in antiferromagnetic insulating systems.
%%%%%%%%%%%%%%%%%%%

\begin{acknowledgements}
We acknowledge support by Deutsche Forschungsgemeinschaft (DFG) under Germany’s Excellence Strategy EXC2181/1-390900948 (the Heidelberg STRUCTURES Excellence Cluster). JA acknowledges support by the IMPRS-QD Heidelberg. KYC was supported by the National Research Foundation (NRF) of Korea (Grant Nos.~2020R1A5A1016518 and RS-2023-00209121). R.S.~acknowledges the financial support provided by the Ministry of Science and Technology in Taiwan under project numbers NSTC 111-2124-M-001-007, Financial support from the Center of Atomic Initiative for New
Materials (AI-Mat), (Project No.~108L9008) and Academia Sinica for the budget of AS-iMATE-113-12.
\end{acknowledgements}

\bibliography{NCTO_QCEP2_arxiv.bbl}

\clearpage
\newpage

\beginsupplement
\onecolumngrid

\section*{\large{Supplemental material: }}

\begin{figure}[!h]
    \centering
    \includegraphics[width=0.7\linewidth]{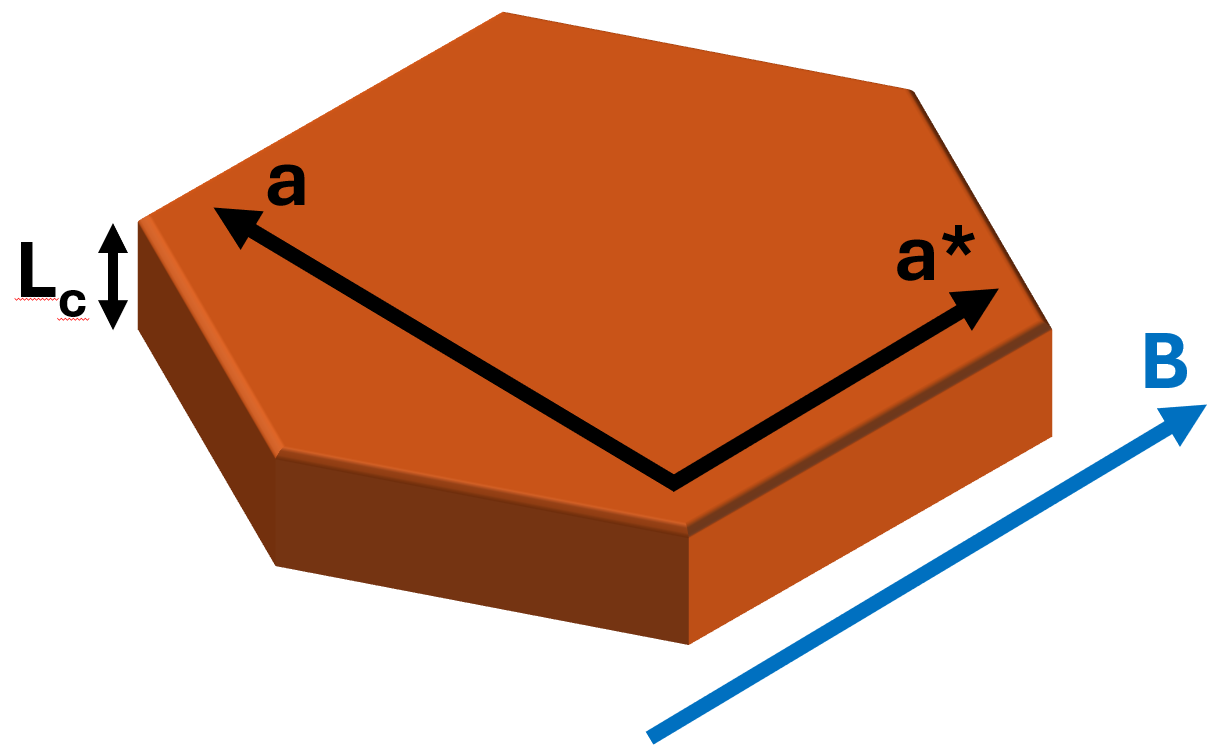}
    \caption{Schematic depiction of the used measurement geometry. The hexagonal shape reflects a stacking of honeycomb layers. All measurements have been performed by measuring the out-of-plane length $L_c$ while the external magnetic field is applied along the Co-Co-bonds within the lattice planes.}
    \label{fig:setup}
\end{figure}

\begin{figure}[!h]
    \centering
    \includegraphics[width = 0.75\textwidth]{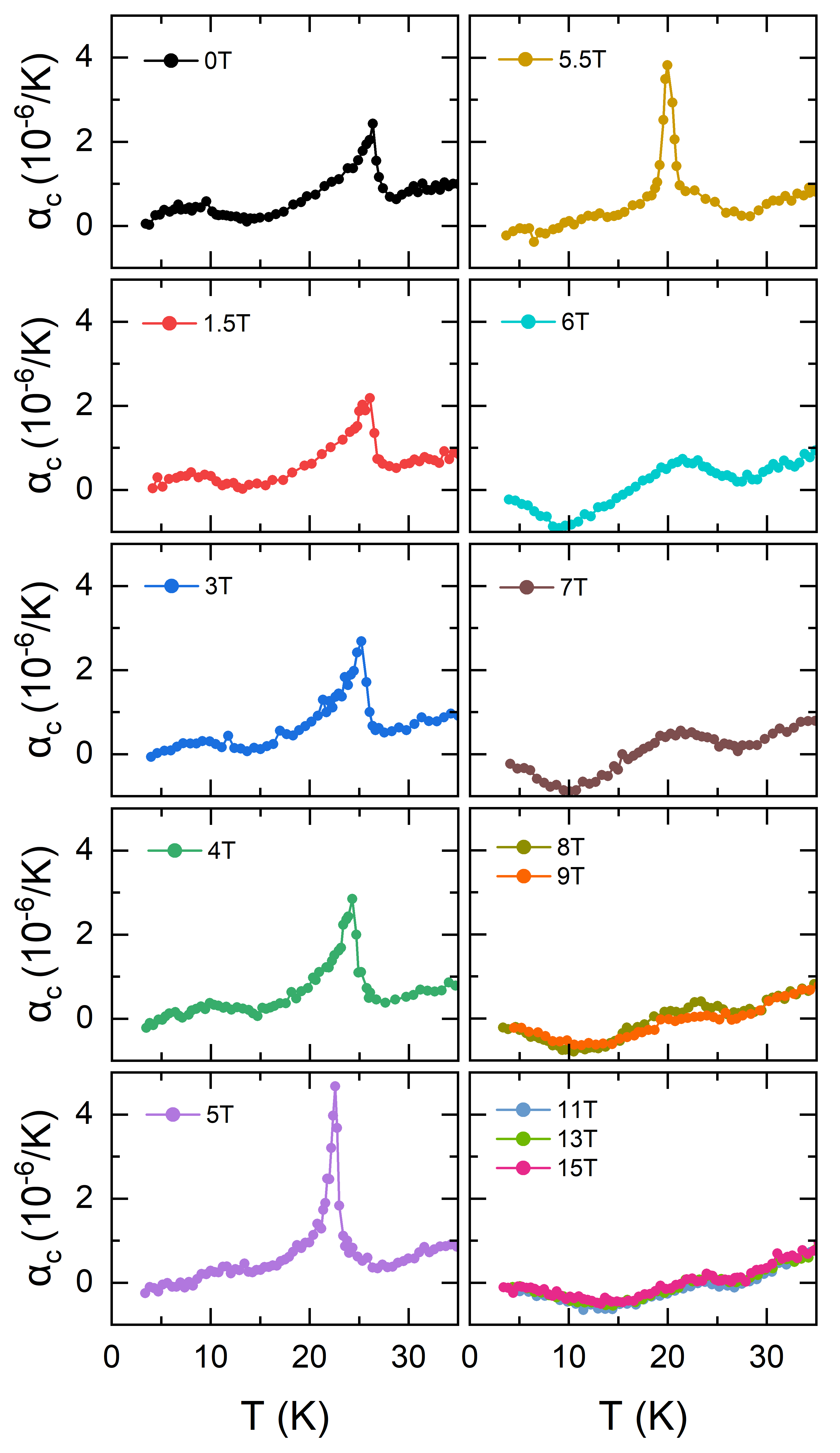}
    \caption{Linear thermal expansion coefficient $\alpha_\mathrm{c}$ for various magnetic fields applied $B||a^*$. }
    \label{fig:TEsupp}
\end{figure}

\begin{figure}[!h]
    \centering
    \includegraphics[width = 0.75\textwidth]{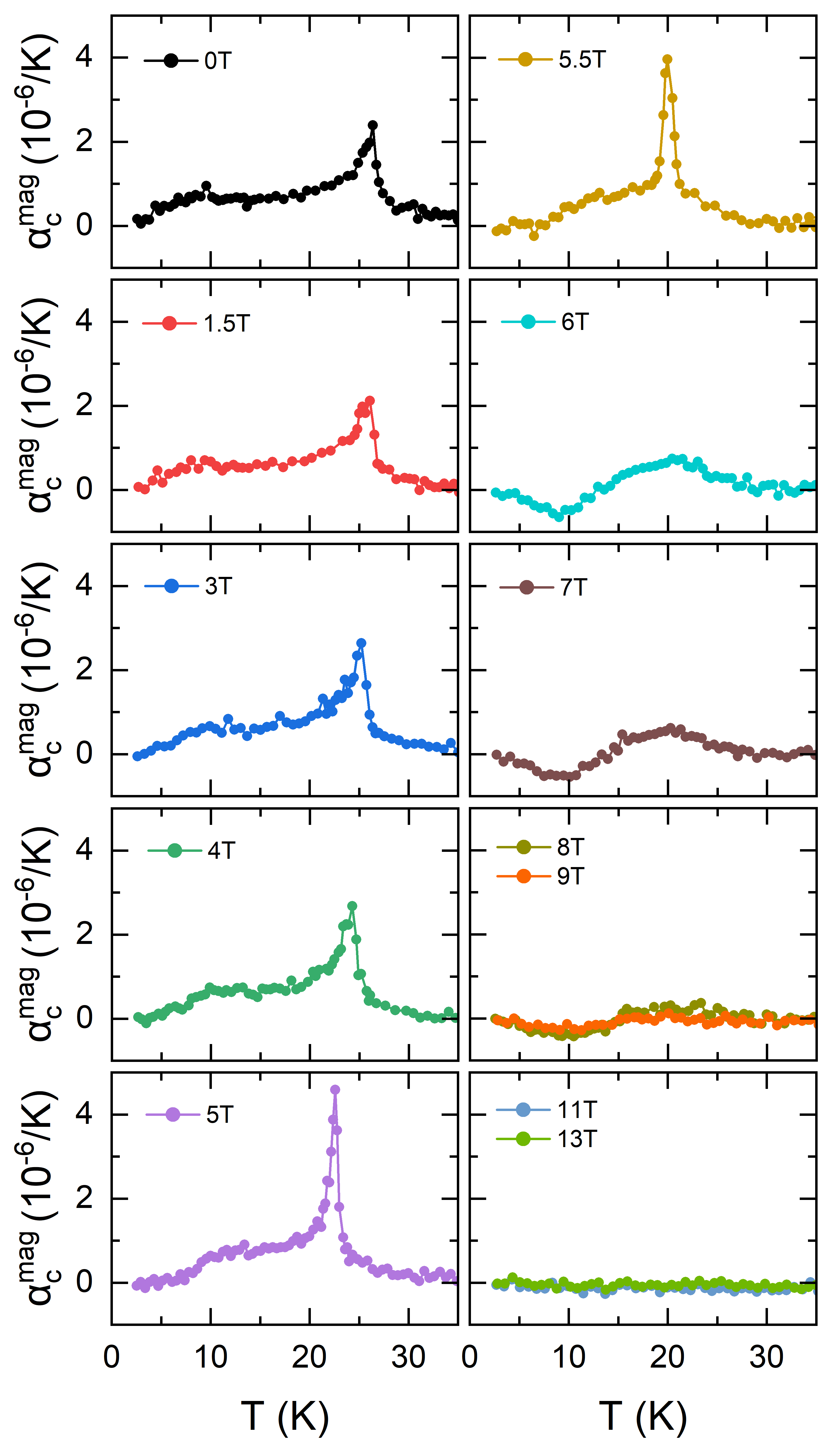}
    \caption{Magnetic contribution $\alpha^{\rm mag}_{\rm c}=\alpha_{\rm c}(B||a^*)-\alpha_{\rm c}(B||a^*=15)$~T at selected magnetic fields applied along the $a^*$ axis. }
    \label{fig:TEmag_supp}
\end{figure}

\begin{figure}[!h]
    \centering
    \includegraphics[width = 0.66\textwidth]{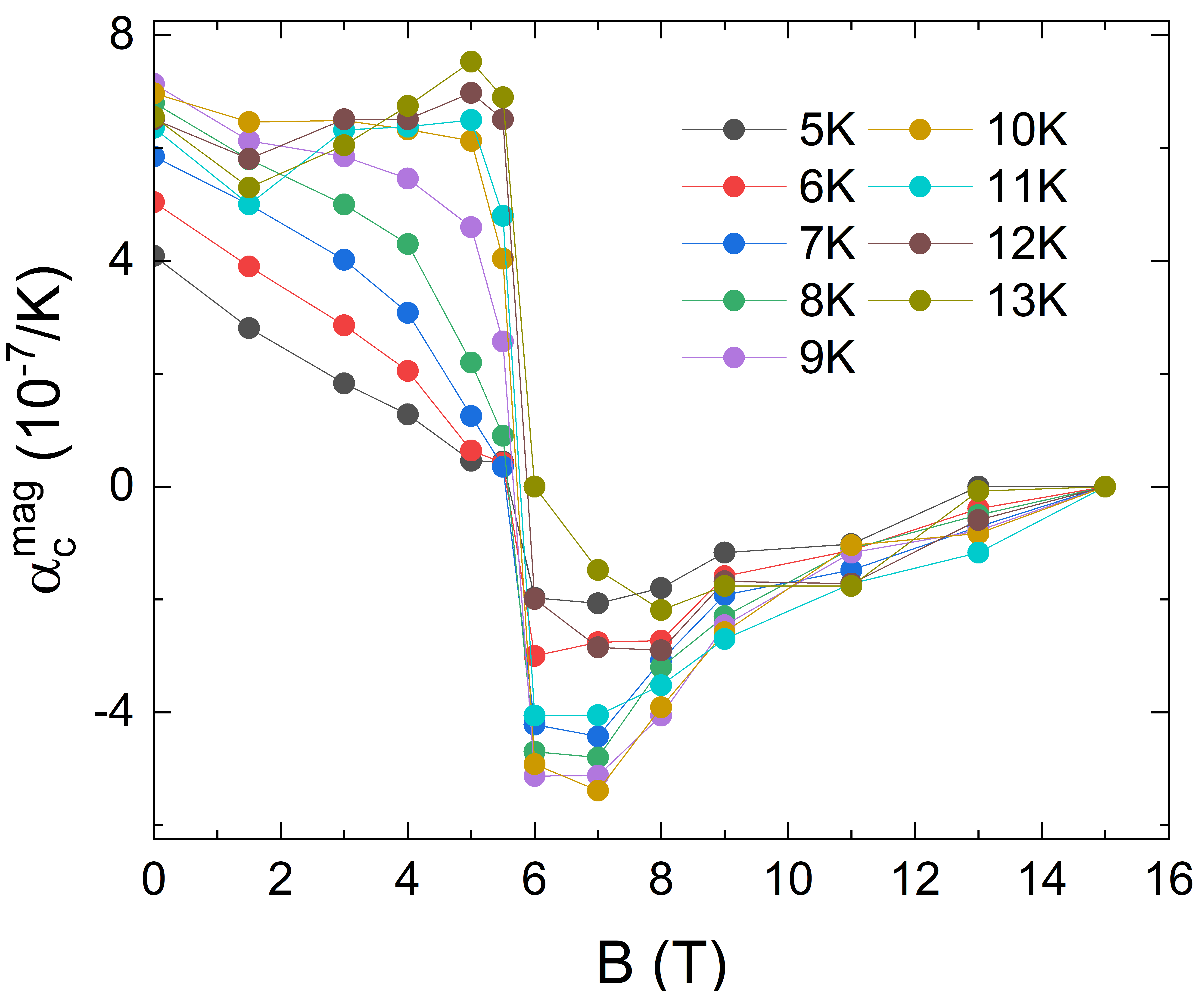}
    \caption{Magnetic contribution to the linear thermal expansion coefficient $\alpha^{\rm mag}_{\rm c}=\alpha_{\rm c}(B||a^*)-\alpha_{\rm c}(B||a^*=15)$~T as a function of magnetic field applied along the $a^*$ axis at selected temperatures.}
    \label{fig:QCEP_diff_T}
\end{figure}

\begin{figure}[!h]
    \centering
    \includegraphics[width = 0.66\textwidth]{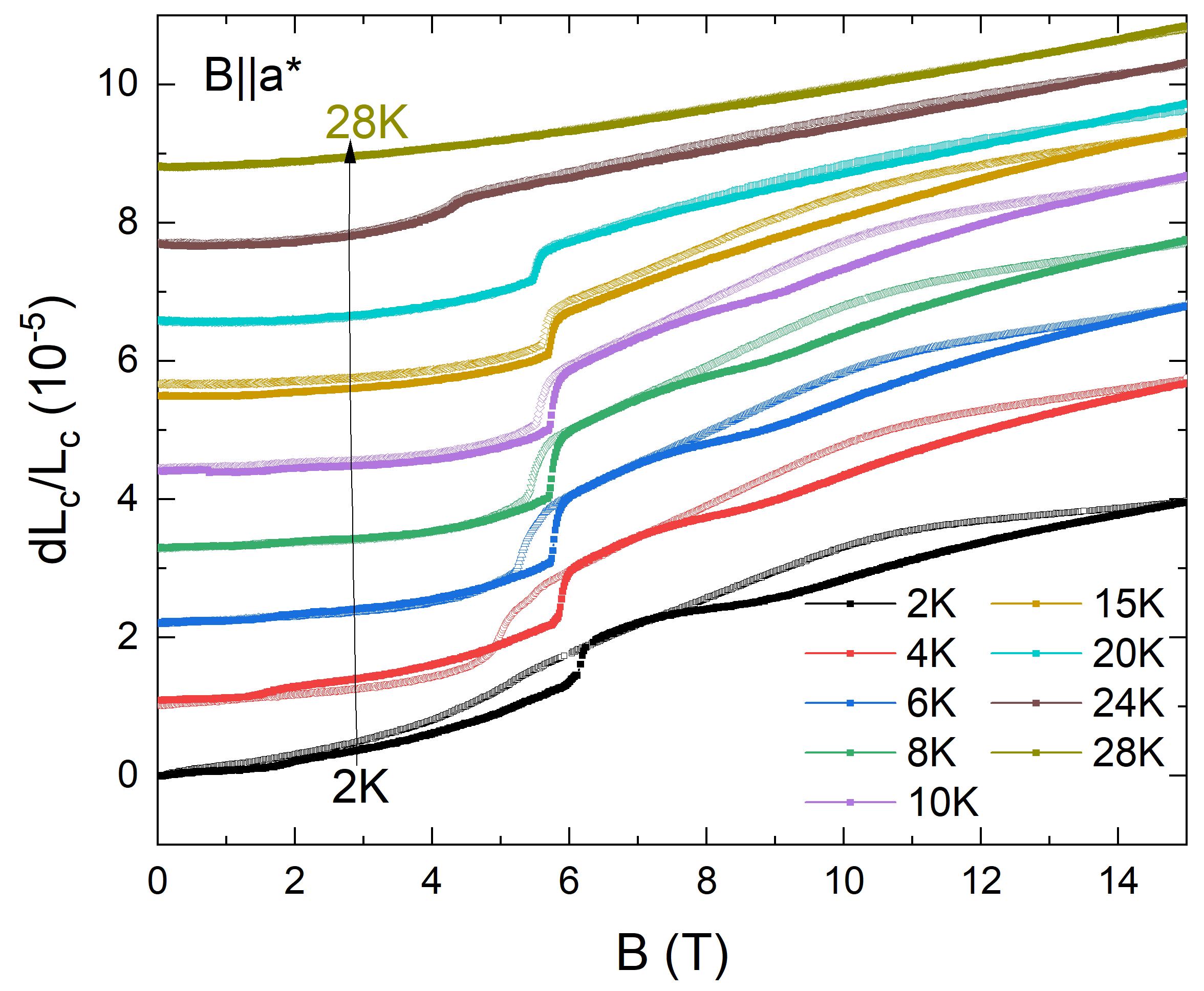}
    \caption{Out-of-plane magnetostriction for $B||a^*$ at selected temperatures. The data are shifted by $1.1 \times 10^{-5}$ along the ordinate.}
    \label{fig:MS_aprime_Tdep}
\end{figure}

\begin{figure}[!h]
    \centering
    \includegraphics[width = 0.65\textwidth]{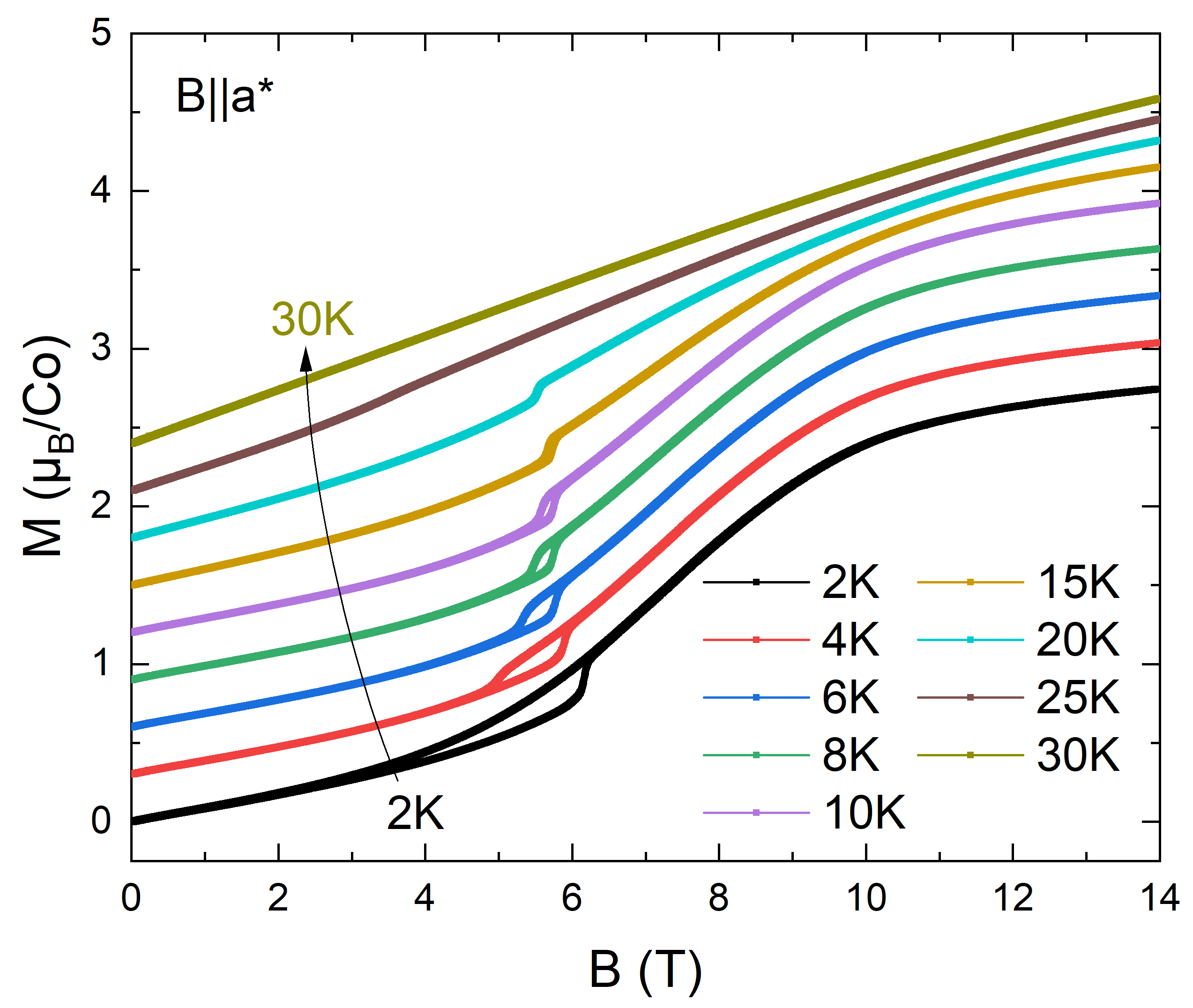}
    \caption{Isothermal magnetization for $B||a^*$ at selected temperatures. The data are shifted by $0.3\,\mathrm{\mu_B/Co}$ along the ordinate.}
    \label{fig:MB_aprime_Tdep}
\end{figure}

\begin{figure}[!h]
    \centering
    \includegraphics[width = 0.64\textwidth]{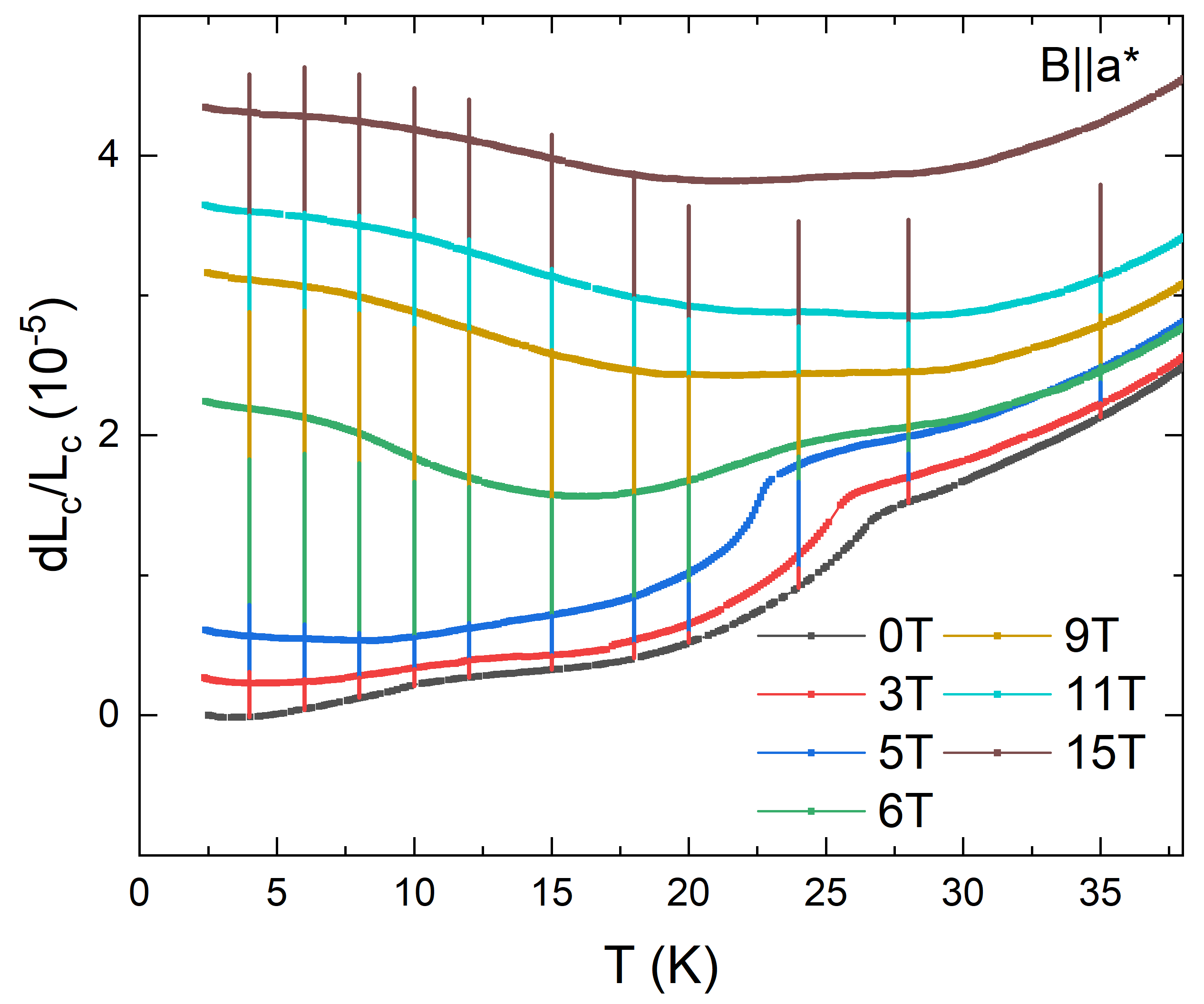}
    \caption{Out-of-plane relative length changes as a function of temperature at selected magnetic fields applied along the $a^*$ axis. The curves are shifted with respect to the magnetostriction data (vertical lines) at $18$~K.}
    \label{fig:TE_infield}
\end{figure}

\begin{figure}[!h]
    \centering
    \includegraphics[width = 0.7\textwidth]{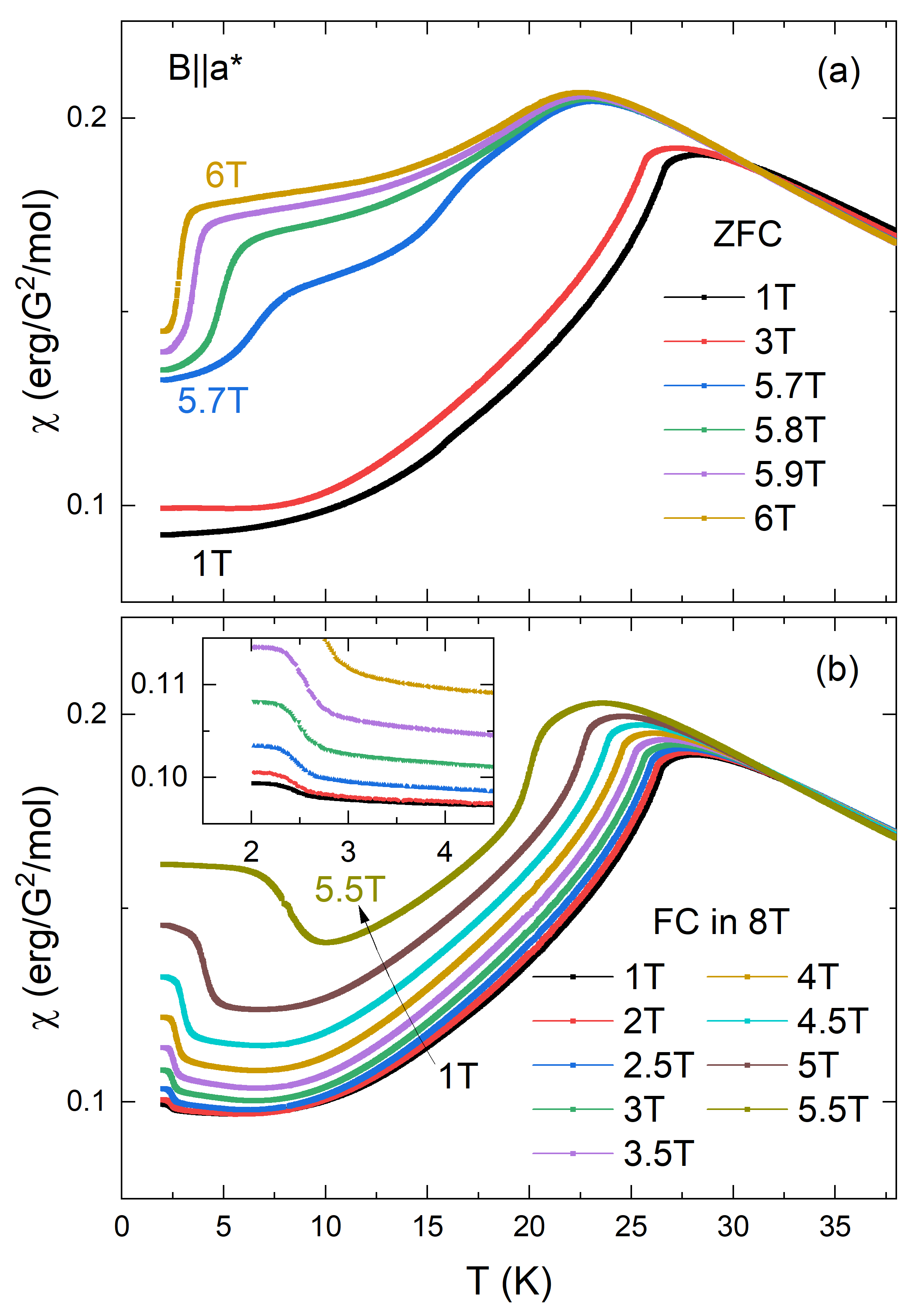}
    \caption{Temperature dependence of the static magnetic susceptibility at selected magnetic fields applied along the $a^*$ axis. In (a) the data is acquired upon heating under zero-field-cooled (ZFC) conditions, i.e., subsequent to cooling to $1.8$~K without external magnetic field and, thus, initially preparing the system in the AFM state. In (b) the data is acquired upon heating subsequent to cooling to $1.8$~K in an external magnetic field of $B||a^* = 8$~T. Hence, the system is initially prepared in the SR state. The inset depicts a zoom into the low-temperature region of the data.}
    \label{fig:MvT_hysteresis}
\end{figure}

\begin{figure}[!h]
    \centering
    \includegraphics[width = 0.7\textwidth]{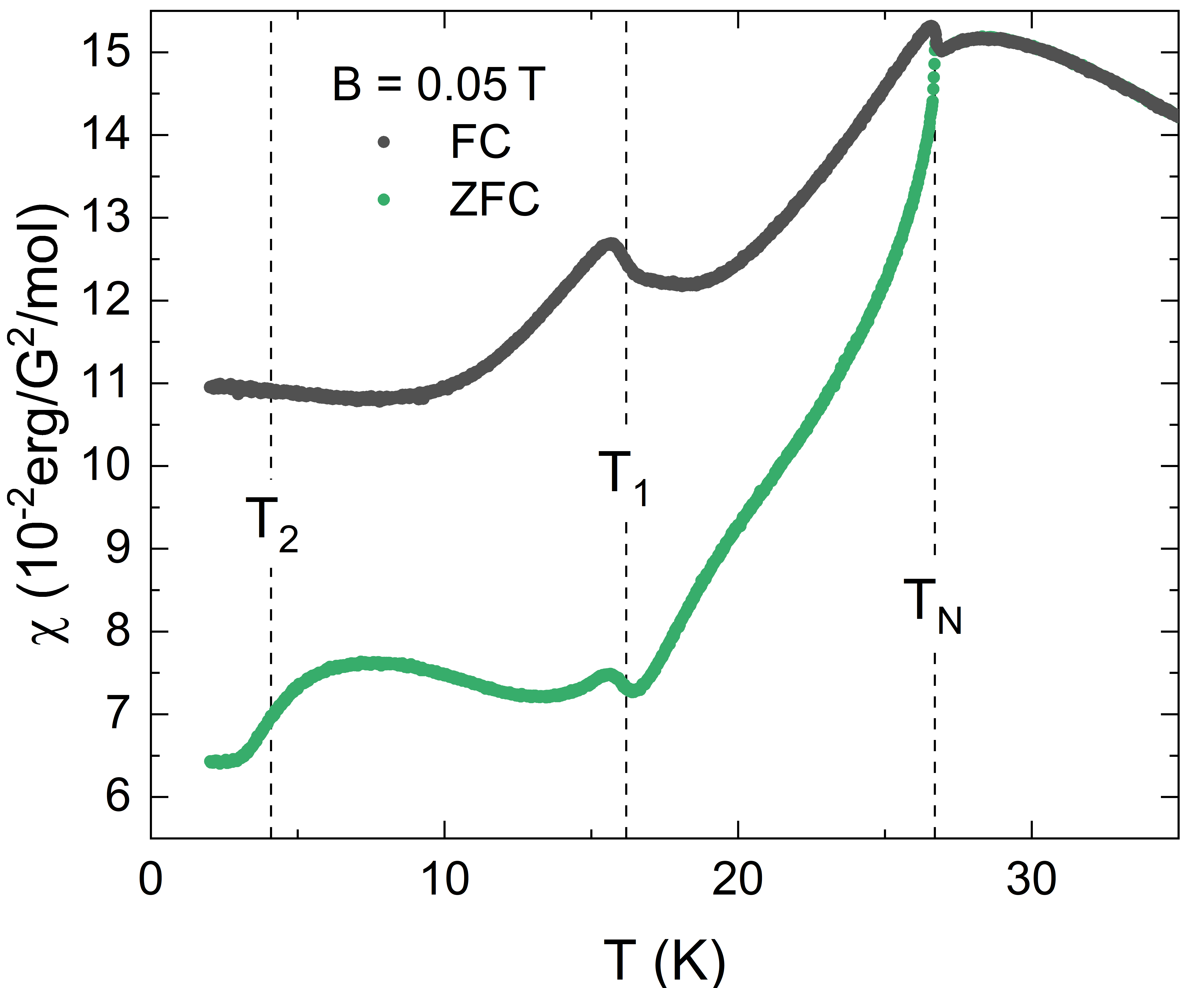}
    \caption{Static magnetic susceptibility at $B = 0.05$~T applied along the $a^*$ axis measured under field-cooled (black) and zero-field-cooled (green) conditions. Dashed vertical lines mark anomalies in the data at $T_\mathrm{N}$, $T_1$ and $T_2$.}
    \label{fig:MvT_lowfields}
\end{figure}

\end{document}